\pdfoutput=1
\documentclass[sort&compress, preprint, review, 12pt]{elsarticle}

\usepackage{graphicx}

\usepackage{amsmath,amssymb}

\usepackage{natbib}
\usepackage[linkcolor=blue, citecolor=blue, colorlinks=true]{hyperref}
\usepackage{booktabs}

\providecommand{\e}[1]{\ensuremath{\times 10^{#1}}}

\journal{Combustion and Flame}

\begin{document}

\begin{frontmatter}



\title{Skeletal mechanism generation for surrogate fuels using directed relation graph with error propagation and sensitivity analysis}


\author[cwru]{Kyle~E. Niemeyer}

\author[uconn]{Chih-Jen Sung\corref{cor1}}
\ead{cjsung@engr.uconn.edu}

\author[GM]{Mandhapati~P. Raju}

\address[cwru]{Department of Mechanical and Aerospace Engineering\\
	Case Western Reserve University, Cleveland, OH 44106, USA}
	
\address[uconn]{Department of Mechanical Engineering\\
	University of Connecticut, Storrs, CT 06269, USA}
	
\address[GM]{General Motors R\&D Tech Center \\
	Warren, MI 48090, USA}
	
\cortext[cor1]{Corresponding author}

\begin{abstract}

A novel implementation for the skeletal reduction of large detailed reaction mechanisms using the directed relation graph with error propagation and sensitivity analysis (DRGEPSA) is developed and presented with examples for three hydrocarbon components, \emph{n}-heptane, \emph{iso}-octane, and \emph{n}-decane, relevant to surrogate fuel development.  DRGEPSA integrates two previously developed methods, directed relation graph-aided sensitivity analysis (DRGASA) and directed relation graph with error propagation (DRGEP), by first applying DRGEP to efficiently remove many unimportant species prior to sensitivity analysis to further remove unimportant species, producing an optimally small skeletal mechanism for a given error limit.  It is illustrated that the combination of the DRGEP and DRGASA methods allows the DRGEPSA approach to overcome the weaknesses of each, specifically that DRGEP cannot identify all unimportant species and that DRGASA shields unimportant species from removal.  Skeletal mechanisms for \emph{n}-heptane and \emph{iso}-octane generated using the DRGEP, DRGASA, and DRGEPSA methods are presented and compared to illustrate the improvement of DRGEPSA.  From a detailed reaction mechanism for \emph{n}-alkanes covering \emph{n}-octane to \emph{n}-hexadecane with 2115 species and 8157 reactions, two skeletal mechanisms for \emph{n}-decane generated using DRGEPSA, one covering a comprehensive range of temperature, pressure, and equivalence ratio conditions for autoignition and the other limited to high temperatures, are presented and validated.  The comprehensive skeletal mechanism consists of 202 species and 846 reactions and the high-temperature skeletal mechanism consists of 51 species and 256 reactions.  Both mechanisms are further demonstrated to well reproduce the results of the detailed mechanism in perfectly-stirred reactor and laminar flame simulations over a wide range of conditions.  The comprehensive and high-temperature \emph{n}-decane skeletal mechanisms are included as supplementary material with this article.

\end{abstract}


\end{frontmatter}

\pagebreak


\section{Introduction}
\label{S:intro}

Combustion of hydrocarbon fuels currently provides 85\% of the energy produced in the United States \cite{Basic-Energy-Sciences-Workshop:2006,Law:2007}.  Renewable sources of energy are being pursued to supplement and eventually replace combustion-based sources, but hydrocarbons will remain the major component for the next few decades.  In the current era of increasing environmental awareness and rising fuel costs, there is considerable demand to improve efficiency and reduce emissions of the next generation combustion technology.  Fuel-flexible designs that can use both conventional and alternative fuels are also desired.

Since computational modeling drives the design of engines and combustors for aerospace, transportation, and energy applications, accurate prediction of fuel combustion and pollutant emissions requires comprehensive detailed reaction mechanisms \cite{Westbrook:2005}.  Liquid transportation fuels contain varying blends of many hydrocarbons.  There has been a recent collaborative effort to develop surrogate models to emulate real fuels to accurately predict combustion properties.  Such surrogate models typically contain mixtures of a small number of appropriate liquid hydrocarbons.  However, detailed reaction mechanisms for surrogates of gasoline \cite{Pitz:2007,Battin-Leclerc:2008}, diesel \cite{Farrell:2007,Battin-Leclerc:2008}, and jet fuels \cite{Violi:2002,Colket:2007,Colket:2008} typically contain large numbers of species and reactions.  For instance, a recently developed detailed mechanism for $ \text{C}_8 - \text{C}_{16} $ \emph{n}-alkane hydrocarbons contains 2115 species and 8157 reactions \cite{Westbrook:2009}, while a mechanism for methyl-decanoate, a biodiesel surrogate, contains 2878 species and 8555 reactions  \cite{Herbinet:2008}.  Despite rapid advancements in computing power, it is generally formidable to integrate such detailed reaction mechanisms into large-scale computational simulations in terms of CPU time and memory requirements.  Since the computational cost of chemistry scales by the third power of the number of species in the worst case when factorizing the Jacobian \cite{Lu:2009}, such large sizes pose problems even in zero-dimensional modeling.  In addition, the wide range of time scales (from nanosecond to second) and the nonlinear coupling between species and reactions induces stiffness when governing equations are solved \cite{Lu:2009}.  Due to these computational demands, reduction of large mechanisms is necessary to facilitate practical simulations using realistic chemistry with modern computational tools.

Skeletal reduction is typically the first step of mechanism reduction, where species and reactions deemed negligible to important phenomena over the range of conditions of interest (e.g.,\ pressure, temperature, and equivalence ratio) are removed from the detailed mechanism.  Much effort has been dedicated to the development of effective skeletal reduction techniques, as reviewed by Griffiths \cite{Griffiths:1995}, Tomlin et al.\ \cite{Tomlin:1997}, and Okino and Mavrovouniotis \cite{Okino:1998}.  Classical skeletal reduction methods include sensitivity analysis \cite{Rabitz:1983,Turanyi:1990a,Turanyi:1990}, principal component analysis \cite{Vajda:1985}, and detailed reduction \cite{Wang:1991}.  Other important methods include lumping \cite{Li:1989,Huang:2005,Pepiot-Desjardins:2008}, genetic algorithms \cite{Edwards:1998,Elliott:2004}, optimization \cite{Bhattacharjee:2003,Oluwole:2006,Mitsos:2008}, and adaptive reduction approaches \cite{Lovas:2002,Schwer:2003,Banerjee:2006,Oluwole:2007,He:2008}.

While mechanism reduction via time scale analysis is a separate approach outside the scope of this paper, such methods can be employed to perform skeletal reduction as well.  Computational singular perturbation (CSP)-based methods \cite{Valorani:2006,Valorani:2007,Prager:2009} analyze the Jacobian matrix to decompose species relations into fast and slow components.  Species are considered important if coupling is strong in either the fast or slow subspace.  However, this approach can overestimate the importance of some species and produce skeletal mechanisms of larger size than other methods \cite{Lu:2006a}.  Another method similar to CSP is level of importance (LOI) analysis \cite{Lovas:2000,Lovas:2002,Lovas:2002a,Lovas:2009}, which combines time scale analysis with sensitivity analysis to rank species importance.  The most recent work \cite{Lovas:2009} using LOI presented skeletal mechanisms for ethylene that are competitive with those generated using other methods \cite{Lu:2005}, though the range of conditions considered in the LOI analysis was much narrower.

The chemistry-guided reduction (CGR) \cite{Zeuch:2008} approach was recently presented and applied to a detailed mechanism for \emph{n}-heptane \cite{Ahmed:2007}.  This method combines lumping and necessity analysis applied to a compact starting mechanism.  The necessity of species is based on reaction-flow analysis toward and from important species.  Though the resulting mechanism sizes are competitive with those from other methods (and the current work), CGR is not explicitly error-controlled and the emphasis on a small starting mechanism could be a possible limitation of the method.

Nagy and Tur\'{a}nyi \cite{Nagy:2009} developed the simulation error minimization connectivity method, based on the original connectivity method proposed by Tur\'{a}nyi \cite{Turanyi:1990}, which exhaustively analyzes sets of important species through Jacobian analysis and selects an optimal mechanism based on an error limit.  The method was shown to provide minimal mechanism sizes for a given error but at a computational expense an order of magnitude above other methods \cite{Nagy:2009}.  This could limit the applicability of the approach to the particularly large mechanisms considered in the current work.

The directed relation graph (DRG) method, originally proposed by Lu and Law \cite{Lu:2005,Lu:2006,Lu:2006a}, recently received significant attention.  This approach uses a directed graph to map the coupling of species and consequently find unimportant species for removal based on selected target species and an acceptable error threshold.  It has been shown to be a particularly efficient and reliable method to reduce large reaction mechanisms \cite{Lu:2006}.  Further development of the DRG method branched into two major directions: (1) DRG-aided sensitivity analysis (DRGASA) \cite{Zheng:2007,Lu:2008}, from the original authors of the DRG method which performs sensitivity analysis on species not removed by DRG to further reduce the mechanism, and (2) DRG with error propagation  (DRGEP) \cite{Pepiot-Desjardins:2008a}, which considers the propagation of error due to species removal down graph pathways.  Another method based on DRG, path flux analysis \cite{Sun:2009}, was recently presented that uses both production and consumption fluxes to define the directed graph and identify important species.  In the current work an approach that integrates the major aspects of DRGEP and DRGASA, DRG with error propagation and sensitivity analysis (DRGEPSA), is presented.  It is illustrated that this combined approach overcomes the weaknesses of the two individual methods.  The DRGEPSA method was initially presented by Raju et al.\ \cite{Raju:2007} and more recently by Niemeyer et al.\ \cite{Niemeyer:2009,Niemeyer:2009a}.  We also note that a similar method combining DRGEP and DRGASA was also recently presented by Zs\'{e}ly et al.\ \cite{Zsely:2009} for the ignition of natural gas mixtures, though not explored in detail as in the current work.

In the following, the methodology and implementation of DRGEPSA for the skeletal reduction of large detailed reaction mechanisms is first discussed in Section \ref{S:method}.  In particular, neat components important to surrogates of gasoline, diesel, and jet fuels are considered.  The weaknesses of DRGEP and DRGASA, and the subsequent improvement of DRGEPSA, are demonstrated with a skeletal reduction of the \emph{n}-heptane detailed mechanism of Curran et al.\ \cite{Curran:1998,Curran:2002} in Section \ref{S:n-hept}.  Additional comparisons are then made in Section \ref{S:iso-oct} using a skeletal reduction of the \emph{iso}-octane detailed mechanism of Curran et al.\ \cite{Curran:2002}  A skeletal mechanism for \emph{n}-decane from the detailed mechanism of Westbrook et al.\ \cite{Westbrook:2009} covering a wide range of conditions is presented in Section \ref{S:n-dec}.  In addition, a high-temperature skeletal mechanism is presented to illustrate the capability of the DRGEPSA method for reduction based on a specific range of conditions.  Conclusions based on the various skeletal reductions as well as suggestions for future work are given in Section \ref{S:conclusion}.

\section{Methodology}
\label{S:method}

The current reduction procedure begins with simulations of constant volume autoignition using the detailed reaction mechanism.  The chemical kinetics data are sampled densely during the ignition evolution and used for the subsequent analysis while the ignition delay results are used to assess the overall performance of the resulting skeletal mechanism.  The DRGEPSA formulation is integrated into the Mechanism Automatic Reduction Software (MARS) implementation \cite{Niemeyer:2010}, which provides a framework for combining multiple mechanism reduction methods into an automatic reduction scheme with minimal required user input.

The DRGEP method is first performed iteratively using the error in ignition delay prediction of the initial skeletal mechanism compared to the results using the detailed mechanism.  The threshold used to identify and remove unimportant species is increased until the maximum error in ignition delay prediction for the given conditions reaches a user-specified limit.  In this manner, the algorithm finds a minimally reduced skeletal mechanism with DRGEP.  The remaining species are then divided into two groups: (1) ``limbo'' species for sensitivity analysis and (2) important species for automatic retention.  Sensitivity analysis is performed on the limbo species to further identify unimportant species, which are then removed until the global ignition delay error reaches the user-specified limit.  In all steps of the reduction process, reactions containing removed species are also eliminated from the mechanism. Specifics of each phase of the skeletal reduction and the DRGEPSA implementation are detailed as follows.

\subsection{DRGEP phase}
\label{S:drgep}

The first phase of DRGEPSA is based on the DRGEP of Pepiot-Desjardins and Pitsch \cite{Pepiot-Desjardins:2008a}, which in turn is an extension of the original DRG of Lu and Law \cite{Lu:2005,Lu:2006,Lu:2006a}.  The current DRGEP implementation includes an improved definition of the direct interaction coefficient employed by Pepiot-Desjardins and Pitsch \cite{Pepiot-Desjardins:2008a}, which was motivated by the shortcomings of the original definition \cite{Pepiot:2005} in situations with long chemical paths involving fast modes \cite{Lu:2006a}.  The DRGEP approach uses a directed relation graph to map the coupling of species in a reaction system, where the graph vertices represent species and directed edges between vertices represent the coupling of species.  The dependence of one species on another is based on a contribution to overall production or consumption rate.  Accurate prediction of the production of a species \emph{A} that is strongly dependent on another species \emph{B} requires the presence of species \emph{B} in the reaction mechanism.  This contribution is expressed with the direct interaction coefficient (DIC):
\begin{equation}
	r_{AB} = \frac{ \bigl | \sum_{i = 1}^{n_R} \nu_{A, i} \omega_{i} \delta_{B}^{i} \bigr | } {\max \left( P_A, C_A \right)  },		\label{E:rAB}
\end{equation}
where
\begin{align}
	P_A &= \sum_{i=1}^{n_R} \max \left( 0, \nu_{A,i} \omega_i \right),		\label{E:PA}\\
	C_A &= \sum_{i=1}^{n_R} \max \left( 0, - \nu_{A,i} \omega_i \right),		\label{E:CA}
\end{align}
\begin{equation}
	\delta_{B}^{i} = \begin{cases}
		1,	&\text{if the $i$th elementary reaction} \\
			& \text{involves species B;}\\
		0,	&\text{otherwise,} \end{cases}							\label{E:dBi}
\end{equation}
\emph{A} and \emph{B} represent the species of interest (with dependency in the $A \rightarrow B$ direction), \emph{i} is the \emph{i}th reaction, $\nu_{A, i}$ is the stoichiometric coefficient of species \emph{A} in the \emph{i}th reaction, $\omega_i$ is the overall reaction rate of the \emph{i}th reaction, and $n_R$ is the total number of reactions.  The DIC is an estimate for the error induced in the overall production or consumption rate of species \emph{A} by the removal of species \emph{B}.

\begin{figure}[htbp]
	\centering
	\includegraphics[width=0.5\linewidth]{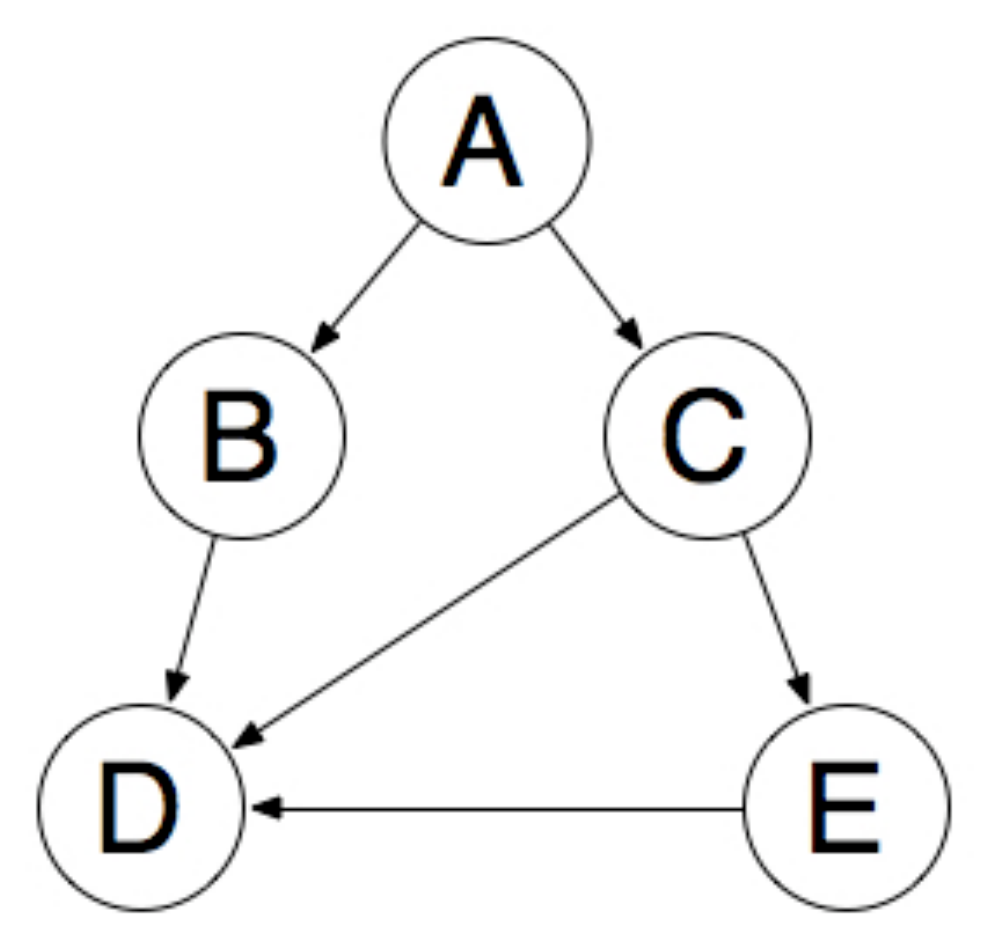}
	\caption{A directed relation graph showing path-dependent species coupling.}
	\label{F:drg-ex}
\end{figure}

After mapping the reaction system, a depth first search is performed starting at user-selected target species (e.g.,\ fuel, oxidizer, important radicals or pollutants) to find the dependency pathways for all species relative to the targets.  A path-dependent interaction coefficient (PIC) represents the error propagation down a certain pathway and is defined as the product of intermediate DICs between the target species \emph{A} and species of interest \emph{B} down a certain path in the directed graph:
\begin{equation} \label{E:rABp}
	r_{AB, p} = \prod_{j = 1}^{n-1} r_{S_j S_{j + 1}}				
\end{equation}
where \emph{n} is the number of species between \emph{A} and \emph{B} in pathway \emph{p} and \emph{S} is a placeholder for the intermediate species starting at species \emph{A} and ending at species \emph{B}.  An overall interaction coefficient (OIC) is then defined as the maximum of all PICs between the targets and each species of interest:
\begin{equation} \label{E:RAB}
	R_{AB} = \max_{\text{all paths $p$}} \left( r_{AB, p} \right)	.	
\end{equation}
For example, Fig.\ \ref{F:drg-ex} shows a simple reaction system where the overall dependence of species \emph{A} on species \emph{D} based on DRG and DRGEP are respectively expressed as:
\begin{equation*}
	R_{AD}^{ \text{DRG} } = \max \left( r_{AB}, r_{BD}, r_{AC}, r_{CD}, r_{CE}, r_{ED} \right)
\end{equation*}
and
\begin{equation*}
	R_{AD}^{ \text{DRGEP} } = \max \left( r_{AD, 1}, r_{AD, 2}, r_{AD, 3} \right),
\end{equation*}
where path one is $A \rightarrow B \rightarrow D$, path two is $A \rightarrow C \rightarrow D$, and path three is $A \rightarrow C \rightarrow E \rightarrow D$ such that
\begin{equation*}
	r_{AD, 1} = r_{AB} \cdot r_{BD}, \quad r_{AD, 2} = r_{AC} \cdot r_{CD}, \quad r_{AD, 3} = r_{AC} \cdot r_{CE} \cdot r_{ED}.
\end{equation*}
  The maximum OIC for each species-target pair is used such that each species is assigned a single OIC for the reaction state currently being considered.

Proper selection of target species for the DRGEP phase is an important consideration.  Unlike in the DRG method, where distance from targets is not taken into account and species only need to be selected to ensure the directed graph is populated by all species \cite{Lu:2005,Lu:2006}, the OIC values used in DRGEP consider species further from targets less important.  Careful selection of target species important to the chemical processes of interest can provide greater reduction in DRGEP by better aligning the OIC values with error in global phenomena (e.g.,\ ignition delay, laminar flame speed).  Liang et al.\ \cite{Liang:2009} performed some analysis of target selection for homogeneous charge compression ignition (HCCI) combustion of \emph{n}-heptane, but the topic merits further study for general applications.  In the current work, using the hydrocarbon parent fuel, oxygen, and nitrogen as targets worked well, though additionally including the hydrogen radical for the \emph{n}-heptane and \emph{iso}-octane reductions resulted in slightly smaller final skeletal mechanisms.

This procedure is performed at each sample point for all kinetics data and the maximum OIC value of all reaction states is used.  This concept can be utilized in dynamic reduction where DRGEP is performed at each grid point and time step.  Liang et al.\ \cite{Liang:2009,Liang:2009a} recently presented such an approach, dynamic adaptive chemistry (DAC), applied to autoignition and HCCI simulations with very good simulation time reduction.  This scheme does not benefit from the sensitivity analysis extension discussed in the current work, however.

The removal of species with OICs below a threshold $ \varepsilon_{EP} $ is considered negligible to the overall production/consumption rates of the target species and therefore such species are unimportant for the given conditions and can be removed from the reaction mechanism.  The optimal threshold is chosen in an iterative manner in this DRGEP implementation.  Using an initially low $ \varepsilon_{EP} $ (e.g.,\ 0.01), a skeletal mechanism is generated and the error in ignition delay prediction (compared to the detailed mechanism) is calculated for all initial conditions using the following:
\begin{equation}
	\delta_{\text{skel}} = \max_{k \in \mathcal{D}} \frac{ \left | \tau_{\text{det}}^k - \tau_{\text{skel}}^k \right | }{ \tau_{ \text{det} }^k } ,
\end{equation}
where $ \tau_{\text{det}}^k $ and $ \tau_{\text{skel}}^k $ are the ignition delay results of the detailed and skeletal mechanisms, respectively, and $ \mathcal{D} $ is the set of autoignition initial conditions.  If the maximum error for this initial skeletal mechanism is above the user-specified error limit, the threshold is decreased.  For this and any subsequent mechanisms, if the maximum error is below the error limit the threshold is increased until the error reaches the specified limit.  This procedure generates a minimal skeletal mechanism using DRGEP for a given error limit before sensitivity analysis is performed.  This routine could stop prematurely, however, because the relationship between the OICs and global error is in general not linear, which will be shown in due course.  A smarter iterative method could produce a better initial mechanism but the sensitivity analysis phase should eliminate any unimportant species missed by DRGEP at a greater computational cost; a better algorithm could make the entire DRGEPSA implementation more efficient.

\subsection{Sensitivity analysis (SA) phase}
\label{S:sa}

The second phase of DRGEPSA is based on the brute-force sensitivity analysis of Zheng et al.\ \cite{Zheng:2007}.  In particular, species with OIC values that satisfy $ \varepsilon_{EP} < R_{AB} < \varepsilon^* $, where $ \varepsilon^* $ is a higher value (e.g.,\ 0.2--0.4), are classified as ``limbo'' species to be analyzed for removal.  Species where $ R_{AB} > \varepsilon^* $ are classified as retained species and are automatically included in the final skeletal mechanism.  This simple sensitivity analysis approach is chosen for computational reasons.  Due to the typically large number of limbo species as well as the large range of autoignition conditions, more sophisticated methods are much more computationally expensive.

Limbo species are first removed from the mechanism one-by-one to find the resulting induced error in ignition delay and then assigned an error measure:
\begin{equation} \label{E:deltaB}
	\delta_B = \left | \delta_{B, \text{ind}} - \delta_{ \text{DRGEP} } \right | ,
\end{equation}
where $ \delta_{B,  \text{ind}} $ is the induced error due to the removal of species \emph{B} with respect to the detailed mechanism and $ \delta_{ \text{DRGEP} } $ is the error of the DRGEP-generated mechanism.  The limbo species are then sorted in ascending order for removal based on $ \delta_B $.  Sorting the species in this manner rather than using the induced error, $ \delta_{B, \text{ind}} $, alone assures the least important species are removed first.  Using induced errors alone for species ranking would not correctly capture the sensitivity of the species with respect to the baseline DRGEP-generated skeletal mechanism.  The removal of species with a $ \delta_{B, \text{ind}} $ above the specified error limit would produce a skeletal mechanism with a maximum error violating the given limit, so such species are removed from the limbo species list and retained in the final mechanism.  The limbo species are then removed from the mechanism in order and the global error is evaluated after each removal.  The skeletal reduction is complete when the maximum error reaches the user-specified error limit.

As shown by Tur\'{a}nyi \cite{Turanyi:1990}, a brute-force sensitivity analysis such as the approach used here does not predict the elimination of species groups.  Removing many low-error species could induce a larger, unpredictable error greater than the error limit and prematurely end the reduction procedure.  To avoid this, the results from sensitivity analysis could be combined with the group-based direct interaction coefficients of the DRGEP implementation of Pepiot-Desjardins and Pitsch \cite{Pepiot-Desjardins:2008a}.  This idea warrants future investigation.

\subsection{MARS}
\label{S:mars}

\begin{figure}[htbp]
	\centering
	\includegraphics[width=0.75\linewidth]{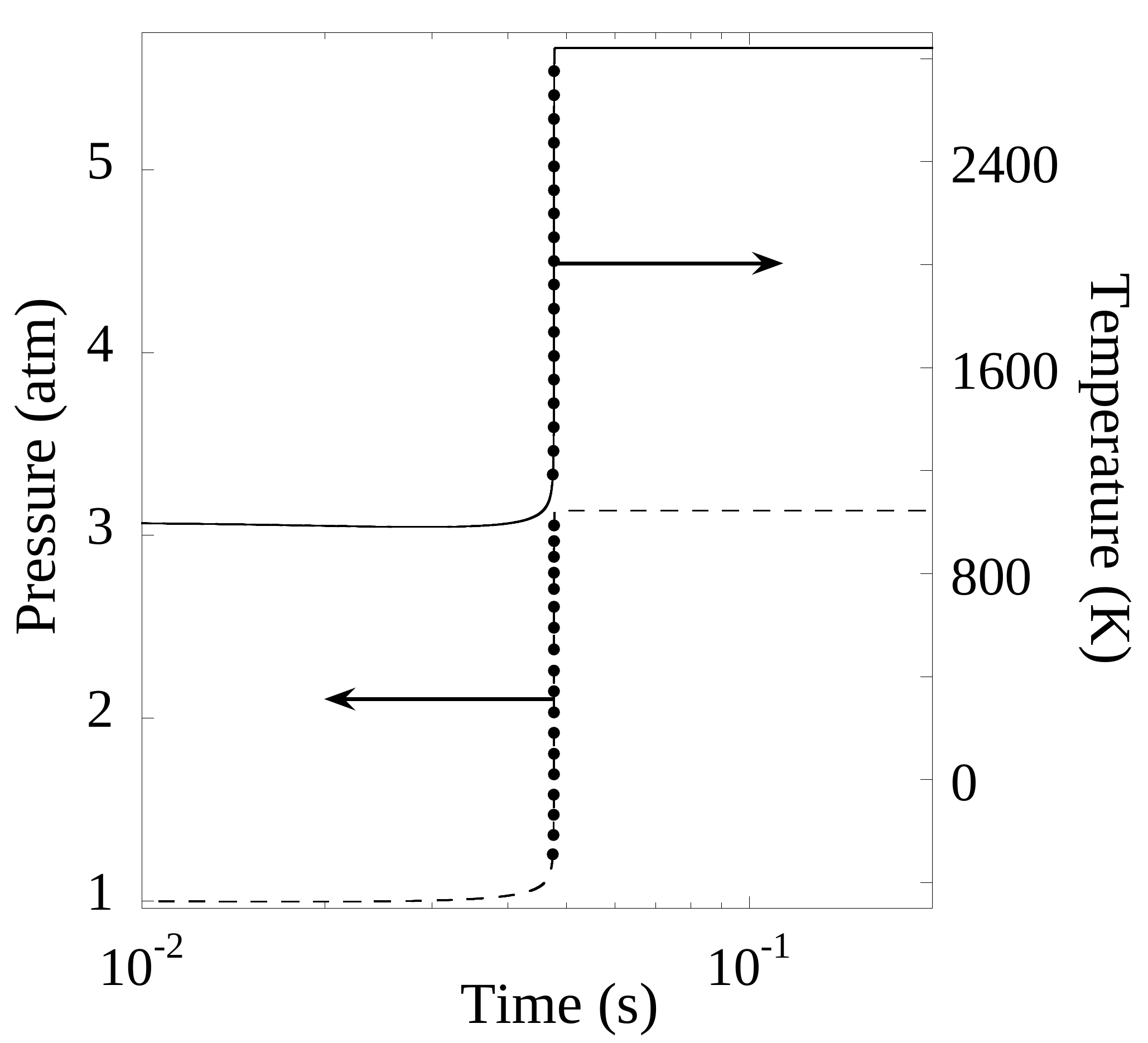}
	\caption{An example of the temperature and pressure data points sampled during the ignition evolution used in the chemical kinetics analysis.}
	\label{F:data-curve}
\end{figure}

The current MARS implementation begins with the use of SENKIN \cite{Lutz:1997} in conjunction with CHEMKIN-III \cite{Kee:1996} to generate numerical solutions of constant volume autoignition using the detailed reaction mechanism over the range of initial conditions for the desired coverage of the skeletal mechanism.  Chemical kinetics data used in the reduction procedure are sampled densely around the ignition evolution, as illustrated in Fig.\ \ref{F:data-curve}.  The ignition delay results from the detailed mechanism are used to measure the error of skeletal mechanisms.  Using a user-defined error limit and the iterative threshold technique described previously, DRGEPSA is used to automatically generate a skeletal mechanism with a minimal number of species.  Similarly, the DRG, DRGASA, and DRGEP methods can be implemented in MARS for purposes of comparison.

\begin{figure}[htbp]
	\centering
	\includegraphics[width=0.75\linewidth]{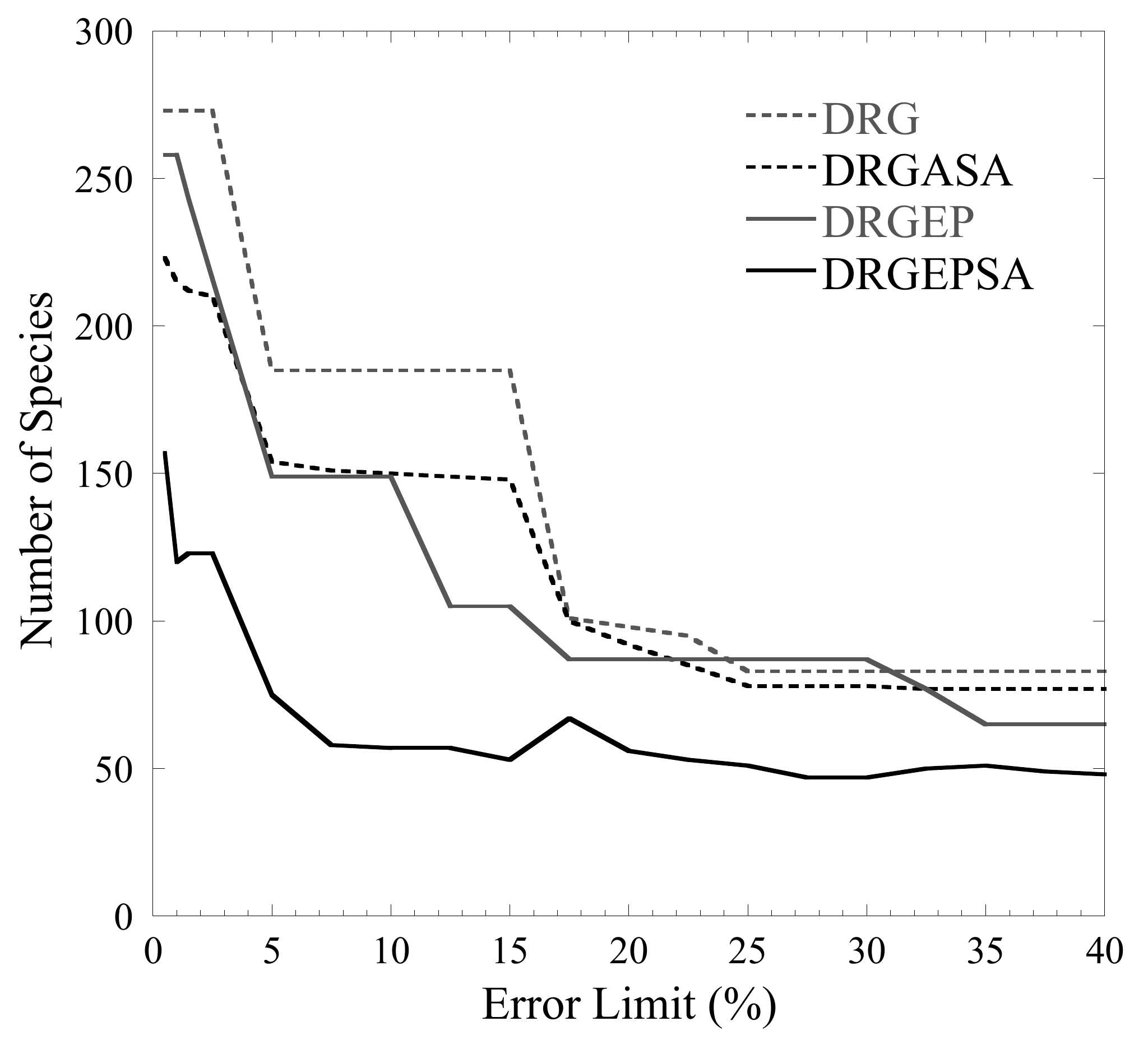}
	\caption{Number of species as a function of error limit for skeletal mechanisms  for \emph{n}-heptane generated using DRG, DRGEP, DRGASA, and DRGEPSA in the MARS implentation.}
	\label{F:error-limit}
\end{figure}

As a demonstration, the DRG, DRGASA, DRGEP, and DRGEPSA methods are implemented in MARS and used to generate skeletal mechanisms for \emph{n}-heptane from the detailed mechanism of \cite{Curran:1998,Curran:2002}, which contains 561 species and 2539 reactions.  For illustration, only three constant volume autoignition initial conditions, at 1000 K, 1 atm, and equivalence ratios of 0.5, 1.0, and 1.5, are used to generate the chemical kinetics data and provide error measures. Error limits from 0.5--40\% are used to compare the methods.  Figure \ref{F:error-limit} shows the numbers of species in the resulting skeletal mechanisms as a function of error limit.  It is clear that DRGEPSA produces smaller size skeletal mechanisms for the full range of error limits, while DRGEP and DRGASA produce mechanisms of similar sizes.  This will be analyzed in greater detail in Section \ref{S:n-hept} at a single error limit for a wider range of initial conditions.  Another interesting trend appears when comparing the skeletal mechanisms resulting from the methods that include sensitivity analysis.  DRGASA and DRGEPSA produce noticeably smaller mechanism sizes even at very strict error limits below 3\%.  This is due to a large number of species with almost negligible induced error that are in general not well identified by DRG or DRGEP.  The reason for this is not yet understood and will be the subject of future study.

\begin{figure}[htbp]
	\centering
	\includegraphics[width=0.75\linewidth]{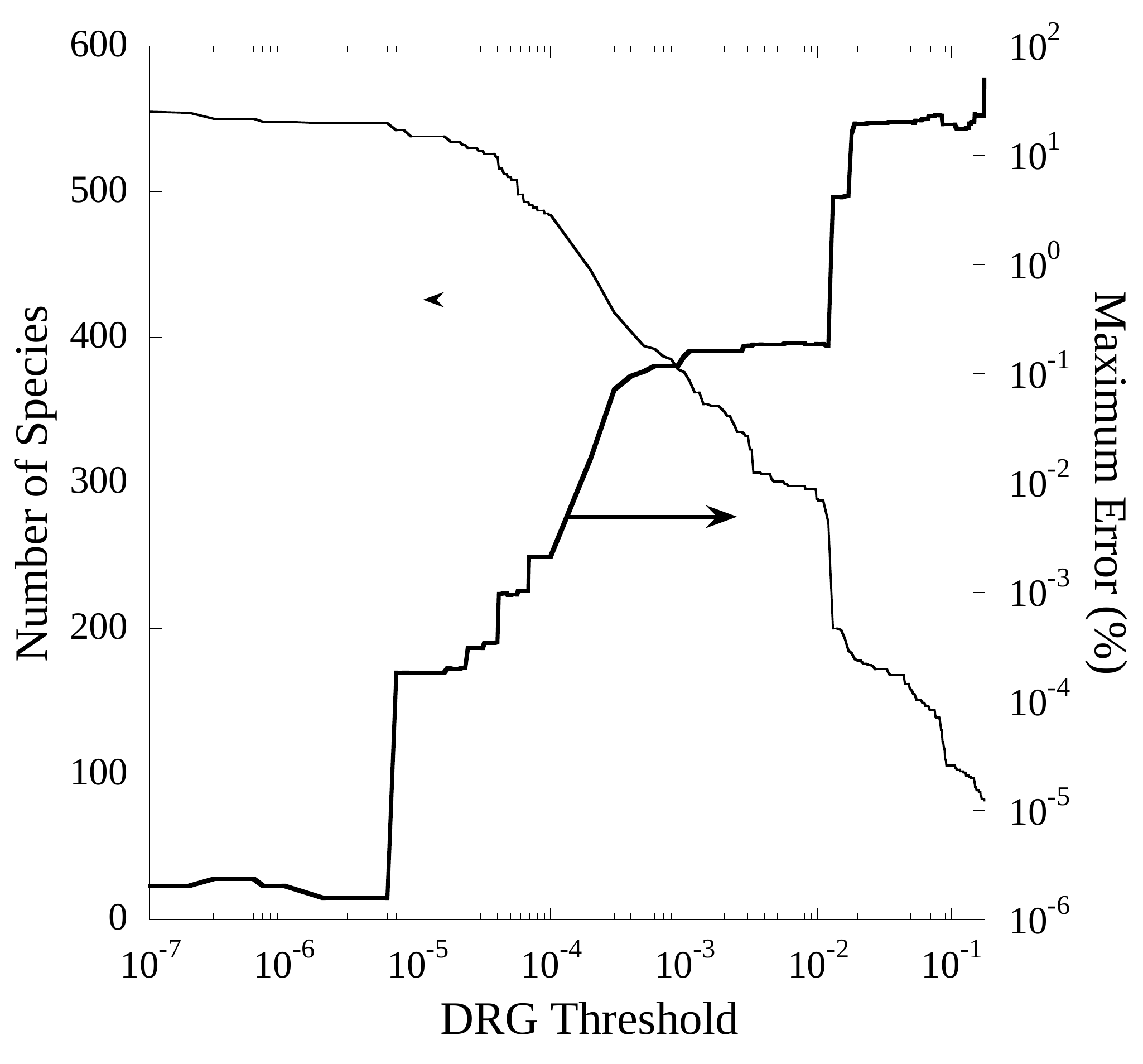}
	\caption{Number of species and maximum error of skeletal mechanisms for \emph{n}-heptane generated by DRG using various threshold values ($ \varepsilon_{DRG} $).}
	\label{F:drg}
\end{figure}

\begin{figure}[htbp]
	\centering
	\includegraphics[width=0.75\linewidth]{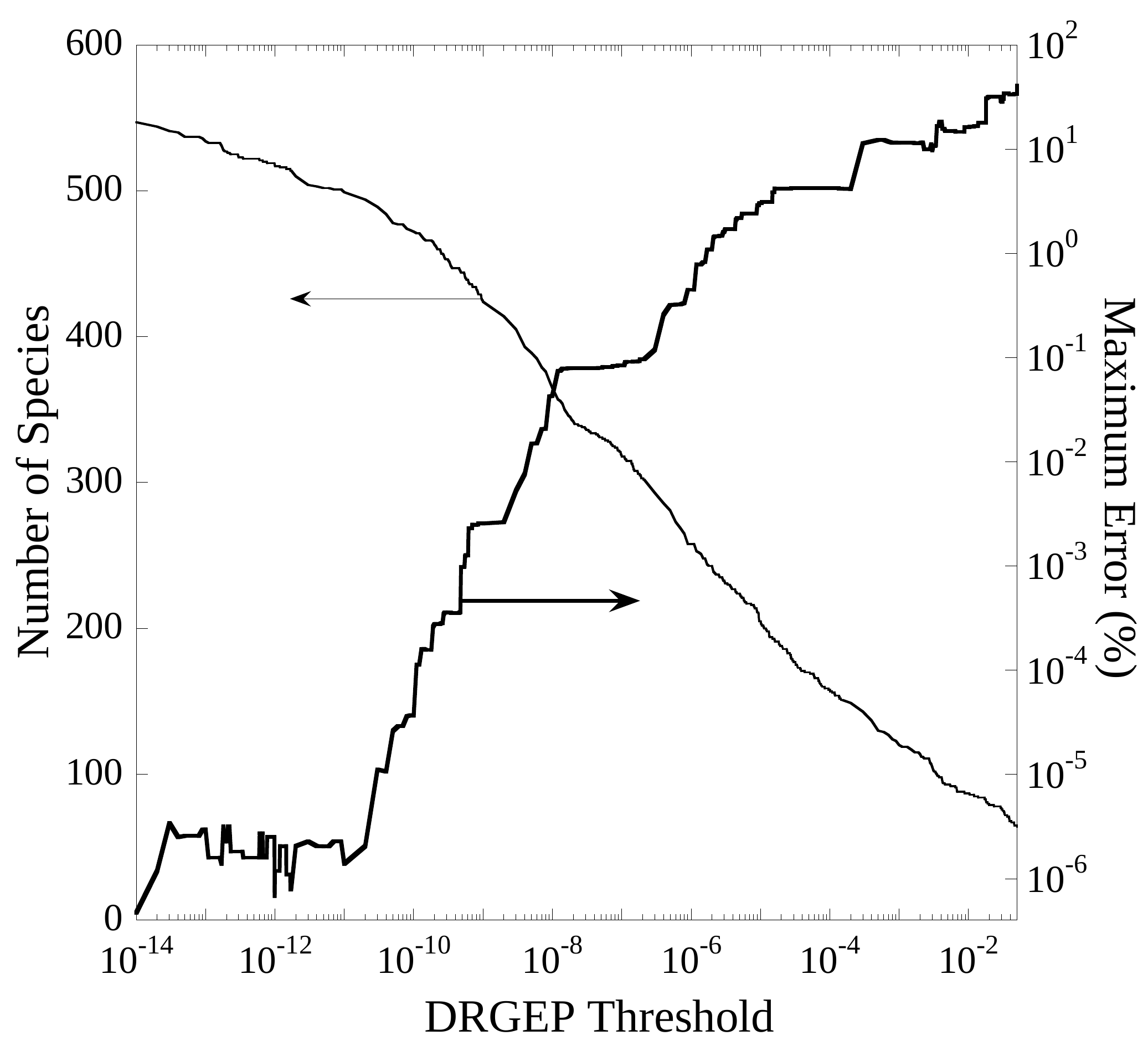}
	\caption{Number of species and maximum error of skeletal mechanisms for \emph{n}-heptane generated by DRGEP using various threshold values ($ \varepsilon_{EP} $).}
	\label{F:drgep}
\end{figure}

In order to analyze DRG and DRGEP in greater detail, Figs.\ \ref{F:drg} and \ref{F:drgep} show number of species and maximum error of skeletal mechanisms generated by DRG and DRGEP, respectively, at various error thresholds ($ \varepsilon_{DRG} $ for DRG and $ \varepsilon_{EP} $ for DRGEP).  As expected, the number of species for skeletal mechanisms produced with both methods monotonically decrease with increasing error threshold.  However, the relationship between the maximum error and error threshold is less clear.  Additionally, the DRG method experiences large jumps in species number and maximum error, which coupled with the iterative threshold selection approach of MARS can explain similar jumps in species number for DRG and DRGASA seen in Fig.\ \ref{F:error-limit}.  DRGEP also experiences similar jumps, though smaller in magnitude.  This implies that the error threshold could be increased to produce slightly smaller skeletal mechanisms than those shown, though only to a certain extent as will be demonstrated in Section \ref{S:n-hept}.

\section{Results and discussion}
\label{S:results}

\subsection{Skeletal reduction of n-heptane}
\label{S:n-hept}

\emph{n}-Heptane is an important primary reference fuel (PRF) for gasoline with a zero octane number and is also important to diesel studies with a cetane number similar to conventional diesel fuel \cite{Curran:1998,Farrell:2007,Pitz:2007}.  Four skeletal mechanisms from the detailed mechanism for \emph{n}-heptane of Curran et al.\ \cite{Curran:1998,Curran:2002}, containing 561 species and 2539 reactions, were generated using DRG, DRGASA, DRGEP, and DRGEPSA to illustrate the individual weaknesses of the DRGASA and DRGEP methods and the subsequent improvement of the combined method.  The DRG and DRGEP methods were performed as the first step of DRGASA and DRGEPSA, respectively.  All the methods used the same autoignition chemical kinetics data and the iterative procedure described in Section \ref{S:drgep} to determine the optimal error threshold values ($ \varepsilon_{DRG} $ for DRG and DRGASA and $ \varepsilon_{EP} $ for DRGEP and DRGEPSA).  The ignition delay error limit was 30\% for the iterative error threshold selection as well as the sensitivity analysis phases of DRGASA and DRGEPSA.  Autoignition chemical kinetics data were sampled from initial conditions covering 600--1600 K, 1--20 atm, and equivalence ratios of 0.5--1.5.  Oxygen, nitrogen, \emph{n}-heptane, and the hydrogen radical were selected as target species for all reduction methods.  As discussed earlier, the hydrogen radical was included to increase the extent of reduction for DRGEP and DRGEPSA for the given error limit.  The DRG and DRGASA results were not affected by this inclusion because the DRG approach does not consider distance from targets in the directed graph and the hydrogen radical was already included in the dependent set.

The skeletal mechanism results for \emph{n}-heptane are shown in Table \ref{T:n-hep1}.  Through the iterative threshold selection procedure, 0.16 was selected as the optimal $ \varepsilon_{DRG} $ to generate a mechanism of 211 species using DRG while 0.01 was selected as the optimal $ \varepsilon_{EP} $ to generate a mechanism of 173 species using DRGEP.  It is seen from Table \ref{T:n-hep1} that the DRGEP and DRGASA methods generate mechanisms of comparable size while DRGEPSA produces a smaller skeletal mechanism for the same given error limit; all methods produce mechanisms of comparable performance.  The mechanism sizes from the various methods are consistent with the earlier trends displayed in Section \ref{S:mars}.

\begin{figure}[htbp]
	\centering
	\includegraphics[width=0.5\linewidth]{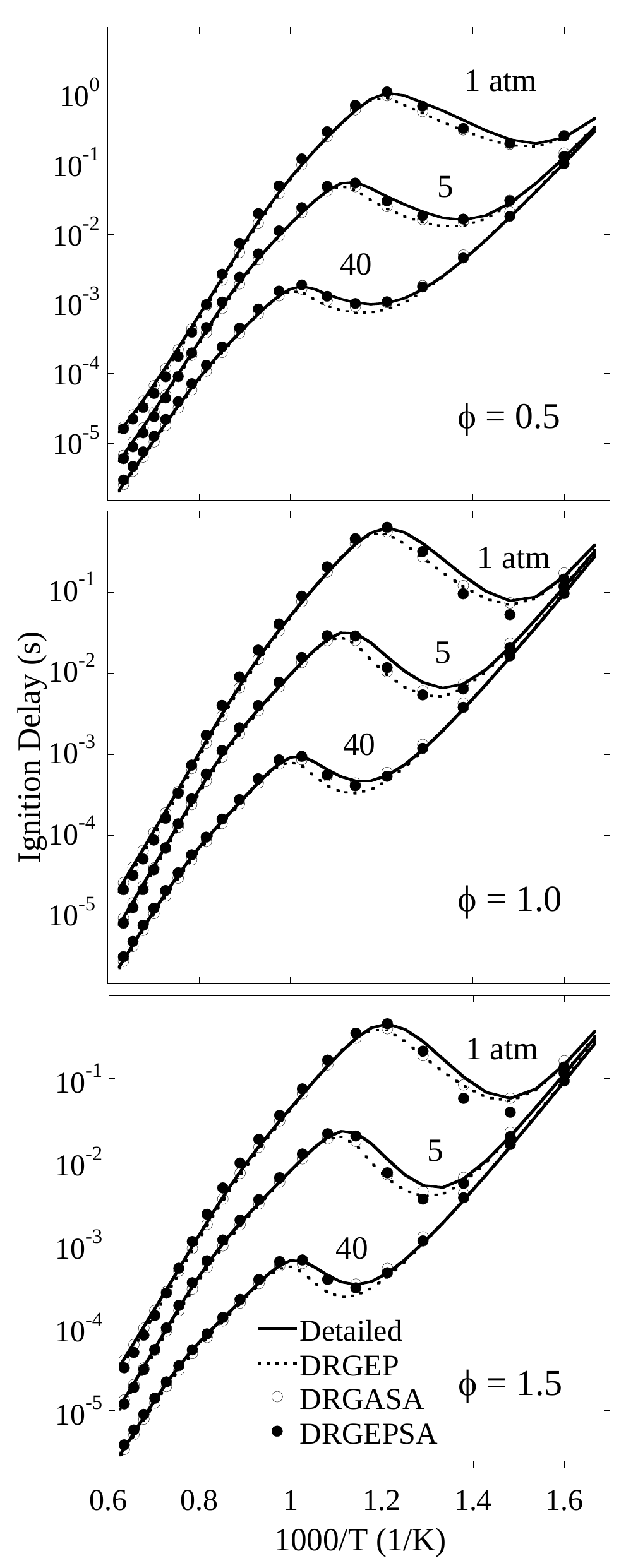}
	\caption{Autoignition validation of \emph{n}-heptane skeletal mechanisms over a range of initial temperatures and pressures, and at varying equivalence ratios.}
	\label{F:nhep-results}
\end{figure}

Validation of the DRGEP, DRGASA, and DRGEPSA skeletal mechanisms was performed and is shown in Fig.\ \ref{F:nhep-results}, covering initial conditions with temperatures of 600--1600 K, pressures of 1, 5, and 40 atm, and equivalence ratios of 0.5, 1.0, and 1.5.  All three skeletal mechanisms exhibit good performance over the full range of conditions, but noticeable discrepancy (limited to 30\% by the reduction procedure) occurs mainly in the negative temperature coefficient (NTC) region.  It is interesting to note that at higher pressures, the larger DRGEP mechanism shows poorer performance in the NTC region than the smaller DRGASA and DRGEPSA mechanisms.

\begin{figure}[htbp]
	\centering
	\includegraphics[width=0.75\linewidth]{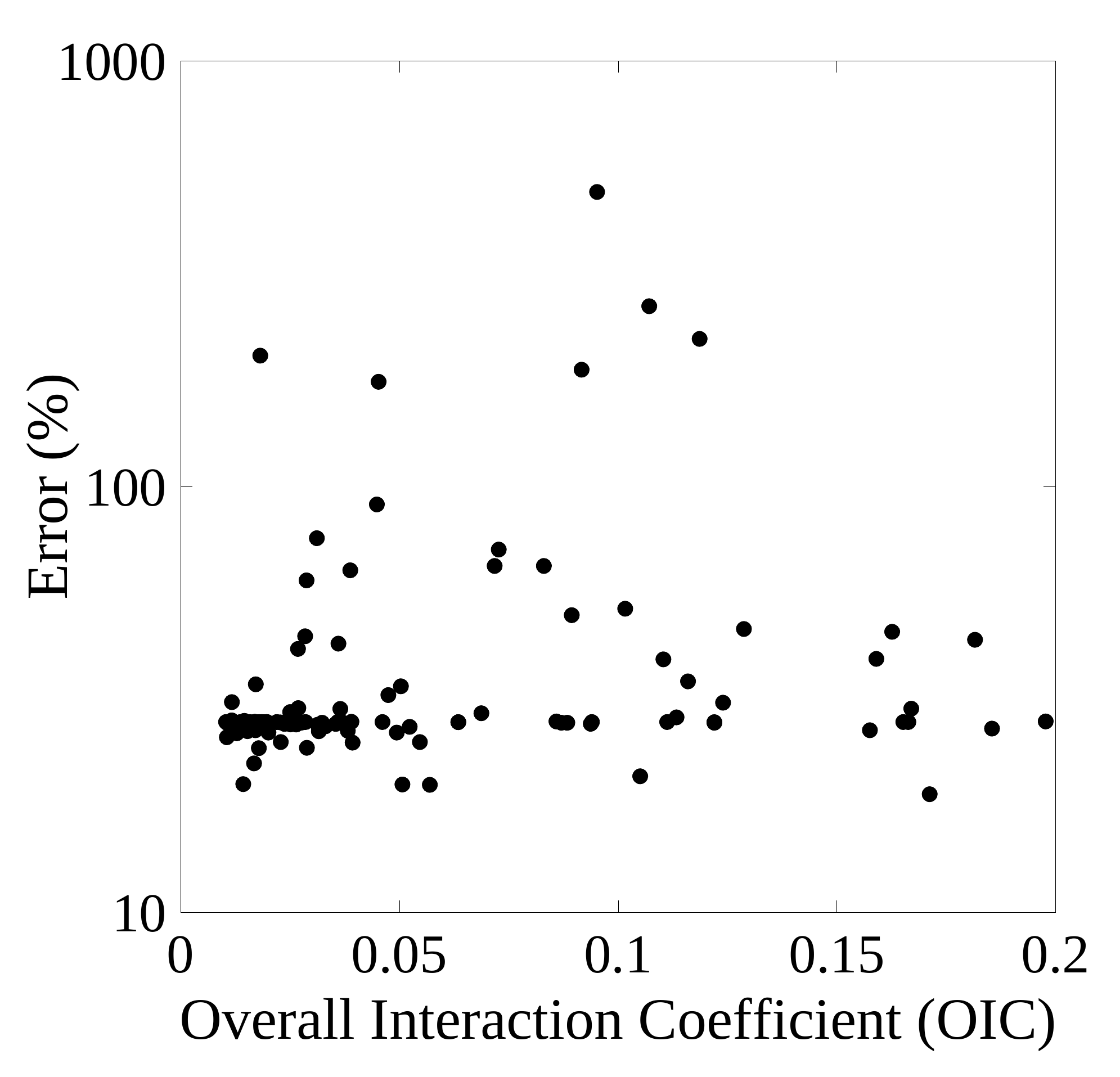}
	\caption{Induced error in ignition delay versus OIC values for species considered with sensitivity analysis phase in DRGEPSA during the \emph{n}-heptane reduction.}
	\label{F:nhep-analysis}
\end{figure}

Figure \ref{F:nhep-analysis} shows the induced errors compared to the OIC values of the species analyzed by sensitivity analysis (limbo species) in the DRGEPSA procedure.  A highly nonlinear relationship is seen, illustrating the need for sensitivity analysis to further reduce the mechanism size.  That is, simply increasing the $ \varepsilon_{EP} $ value would not be sufficient to remove unimportant species in this range of OIC values such that sensitivity analysis is needed following the DRGEP phase to generate a minimal skeletal mechanism.  The OIC well identifies unimportant species but may lose accuracy for species of higher importance and induced error.

The improvement of DRGEPSA over DRGASA is evident by the smaller final mechanism size and equivalent performance; DRGASA performs slightly better than DRGEP here, but the method cannot identify all unimportant species due to species ``shielding.''  This occurs because the DRG phase of DRGASA uses a DIC to rank species \cite{Lu:2005,Lu:2006,Lu:2006a}, which does not consider distance from targets and can inflate species importance such that species are automatically retained rather than assessed with sensitivity analysis.  Table \ref{T:n-hep2} contains the results of the sensitivity analysis phase from DRGASA and DRGEPSA in the current comparison.  46 species out of the 129 species automatically retained by DRGASA were eliminated from the final mechanism by DRGEPSA; specifically, 17 were removed by the DRGEP phase and 29 were removed by the SA phase in DRGEPSA, illustrating the shielding effect in DRGASA.  Although eliminating the shielding increases the extent of total reduction, the greater number of species considered for sensitivity analysis can cause the DRGEPSA reduction to be more computationally expensive.

A skeletal mechanism for \emph{n}-heptane was previously generated with a different implementation of the DRGASA method by Lu and Law \cite{Lu:2006,Lu:2008} using a two-stage DRG followed by sensitivity analysis.  This approach used a similar autoignition initial condition range but also included perfectly-stirred reactor (PSR) kinetics data in the reduction procedure.  The first stage of DRG, using $ \varepsilon_{DRG} = 0.1 $ produced an initial mechanism with 290 species, and the second stage, applying DRG again to the resulting mechanism, used $ \varepsilon_{DRG} = 0.19 $ to produce a mechanism with 188 species and 939 reactions.  Following the sensitivity analysis phase, a final skeletal mechanism with 78 species and 359 reactions was obtained with approximately 30\% maximum error.  While this reduction provided a skeletal mechanism smaller than both the DRGASA and DRGEPSA mechanisms shown here, the purpose of the current work is to compare the DRGEP, DRGASA, and DRGEPSA methods alone rather than specific strategies or implementations of employing such methods.  The methods compared here used the same kinetics data and reduction procedure and differed only in the manner of ranking and selecting species for removal.

\subsection{Skeletal reduction of iso-octane}
\label{S:iso-oct}

\emph{iso}-Octane is the other important PRF for gasoline with 100 on the octane scale \cite{Curran:2002,Pitz:2007}.  Four skeletal mechanisms from the detailed mechanism for \emph{iso}-octane of Curran et al.\ \cite{Curran:2002}, containing 857 species and 3606 reactions, were generated using DRG, DRGASA, DRGEP, and DRGEPSA to illustrate the greater reduction capability of the final method.  All methods used the same constant volume autoignition data sampled from initial conditions ranging over 600--1600 K, 1--20 atm, and equivalence ratios of 0.5--1.5.  The maximum error limit was 30\%.  Oxygen, nitrogen, \emph{iso}-octane, and the hydrogen radical were selected as target species for all methods.

The skeletal reduction results for \emph{iso}-octane are shown in Table \ref{T:iso-oct}.  Through the iterative selection procedure, 0.15 was selected as the optimal $ \varepsilon_{DRG} $ and 0.004 as the optimal $ \varepsilon_{EP} $.  The results for the various mechanism sizes are similar to those from the \emph{n}-heptane reduction; DRGEP and DRGASA produced mechanisms of comparable size and performance while DRGEPSA gives a mechanism substantially smaller than both for similar error.  The mechanism sizes from the various methods are consistent with the earlier trends displayed in Section \ref{S:mars}.

\begin{figure}[htbp]
	\centering
	\includegraphics[width=0.5\linewidth]{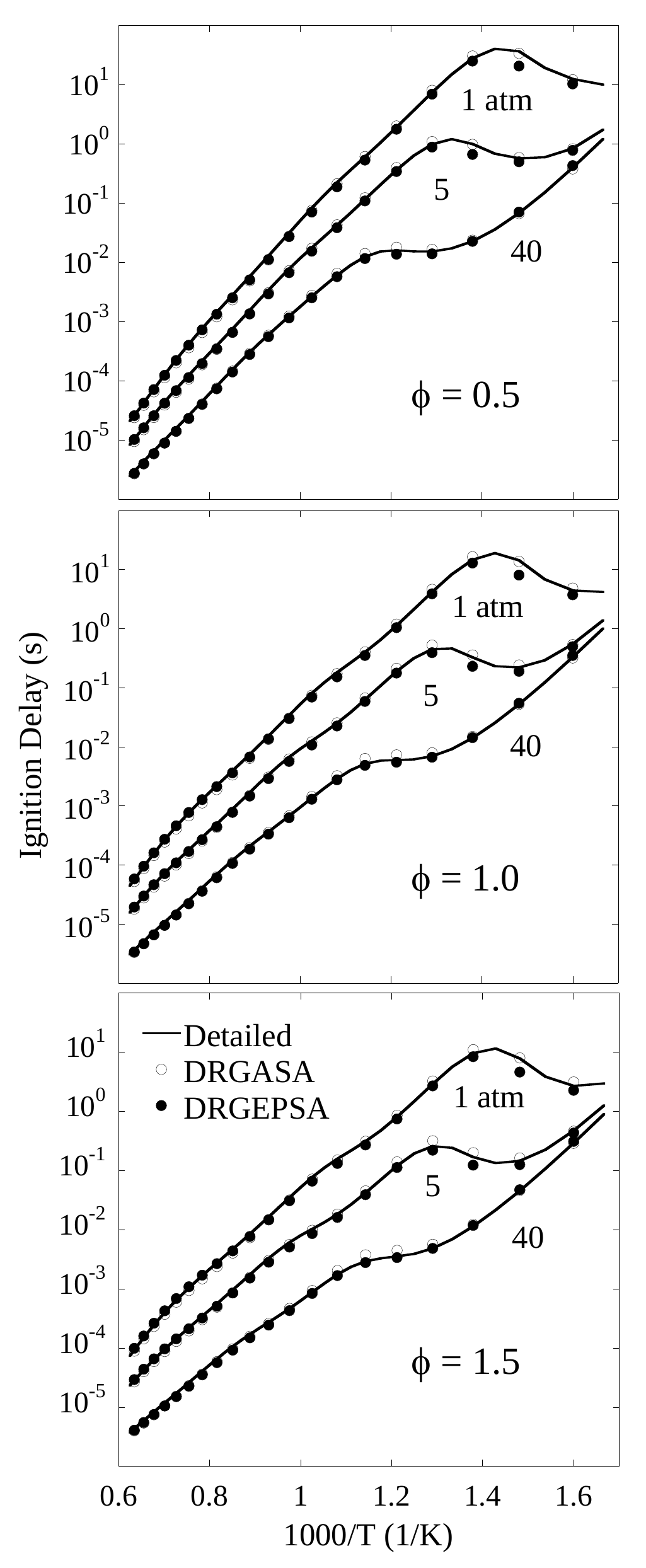}
	\caption{Autoignition validation of \emph{iso}-octane skeletal mechanisms over a range of initial temperatures and pressures, and at varying equivalence ratios.}
	\label{F:isooct-results}
\end{figure}

Validation of the DRGASA and DRGEPSA skeletal mechanisms was performed and is shown in Fig.\ \ref{F:isooct-results}, covering temperatures of 600--1600 K, pressures of 1, 5, and 40 atm, and equivalence ratios of 0.5, 1.0, and 1.5.  Good agreement between the ignition delay predictions of the skeletal mechanisms and the detailed mechanism is observed, with some discrepancy in the NTC region as in the \emph{n}-heptane cases.

Pepiot-Desjardins and Pitsch \cite{Pepiot-Desjardins:2008a} generated skeletal mechanisms for \emph{iso}-octane at various levels of complexity using an implementation of the DRGEP method.  Autoignition chemical kinetics data covering a similar range of conditions to the current study were used, while the targets used were \emph{iso}-octane, CO, $\text{CO}_2$, and temperature through heat release data.  The relevant skeletal mechanism generated using DRGEP consisted of 196 species and 1762 irreversible reactions with a maximum error of 15\%; this is of the same order as the DRGEP mechanism produced in the current study, though slightly smaller.  In addition to a smaller error limit, their DRGEP implementation contains certain extensions not included in the current DRGEP, such as group-based DIC, scaling of DIC based on element flux, and an integrity check.

Similar to the strategy previously adopted for \emph{n}-heptane, Xin et al.\ \cite{Xin:2009} recently presented a skeletal mechanism for \emph{iso}-octane generated with the DRGASA method using a two-stage DRG phase.  The reduction used a similar autoignition initial condition range to that of the current study, but included PSR in addition to autoignition chemical kinetics data.  The first stage of DRG used $ \varepsilon_{DRG} = 0.1 $ to generate an initial mechanism with 347 species and the second stage used $ \varepsilon_{DRG} = 0.17 $ to generate a mechanism with 233 species and 959 reactions.  After the sensitivity analysis phase, the final skeletal mechanism consisted of 112 species and 481 reactions.  As in the \emph{n}-heptane case, the two-stage DRG strategy in conjunction with DRGASA implementation is not being compared in the current work, only the DRGASA and DRGEPSA methods alone. 

\subsection{Skeletal reduction of n-decane}
\label{S:n-dec}

\emph{n}-Decane is an important diesel surrogate component \cite{Farrell:2007} and a primary component for jet fuel surrogates \cite{Colket:2007,Colket:2008}.  Two skeletal mechanisms for \emph{n}-decane were generated using the DRGEPSA method from the detailed mechanism for \emph{n}-alkanes covering \emph{n}-octane through \emph{n}-hexadecane of Westbrook et al.\ \cite{Westbrook:2009} which contains 2115 species and 8157 reactions.  The first skeletal reduction was performed using constant volume autoignition data sampled from initial conditions covering 600--1600 K and 1--20 atm, while the second skeletal reduction was limited to the high-temperature regime (1000--1300 K) and atmospheric pressure; both covered equivalence ratios of 0.5--1.5.  The error limit for both reductions was 30\%; oxygen, nitrogen, and \emph{n}-decane were selected as target species for both reductions.  In this case, selecting the hydrogen radical as an additional target did not increase the extent of reduction for the given error limit.

The DRGEP phase of the comprehensive reduction generated an initial skeletal mechanism with 381 species and 1865 reactions with a maximum error of 27\% using $\varepsilon_{EP}=1.4\e{-3}$ while the high-temperature reduction used $\varepsilon_{EP} =  0.007$ to generate  a mechanism with 68 species and 391 reactions with a maximum error of 24\%.  The final comprehensive skeletal mechanism following sensitivity analysis consists of 202 species and 846 reactions with a maximum error of 25\% and the high-temperature skeletal mechanism consists of 51 species and 256 reactions with a maximum error of 27\%.

\begin{figure}[htbp]
	\centering
	\includegraphics[width=0.5\linewidth]{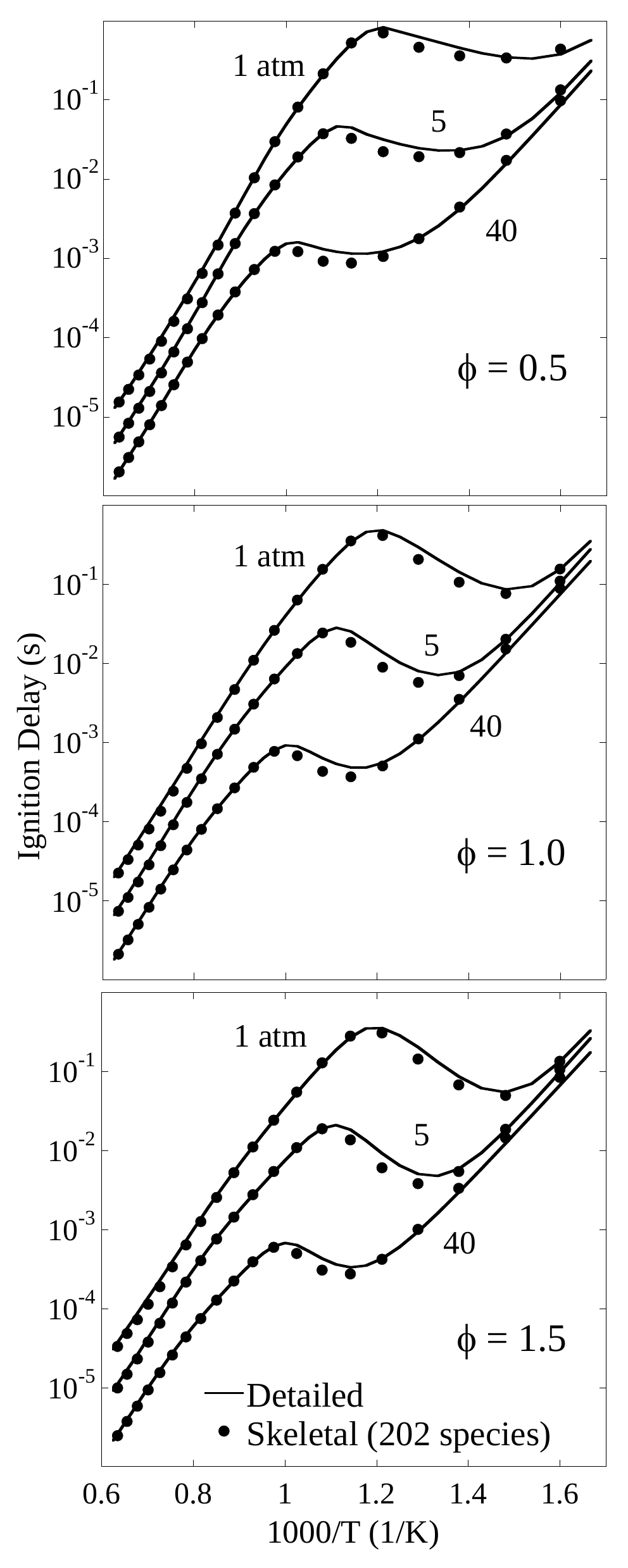}
	\caption{Autoignition validation of comprehensive \emph{n}-decane skeletal mechanism (202 species and 846 reactions) over a range of initial temperatures and pressures, and at varying equivalence ratios.}
	\label{F:ndec-comp-results}
\end{figure}

\begin{figure}[htbp]
	\centering
	\includegraphics[width=0.5\linewidth]{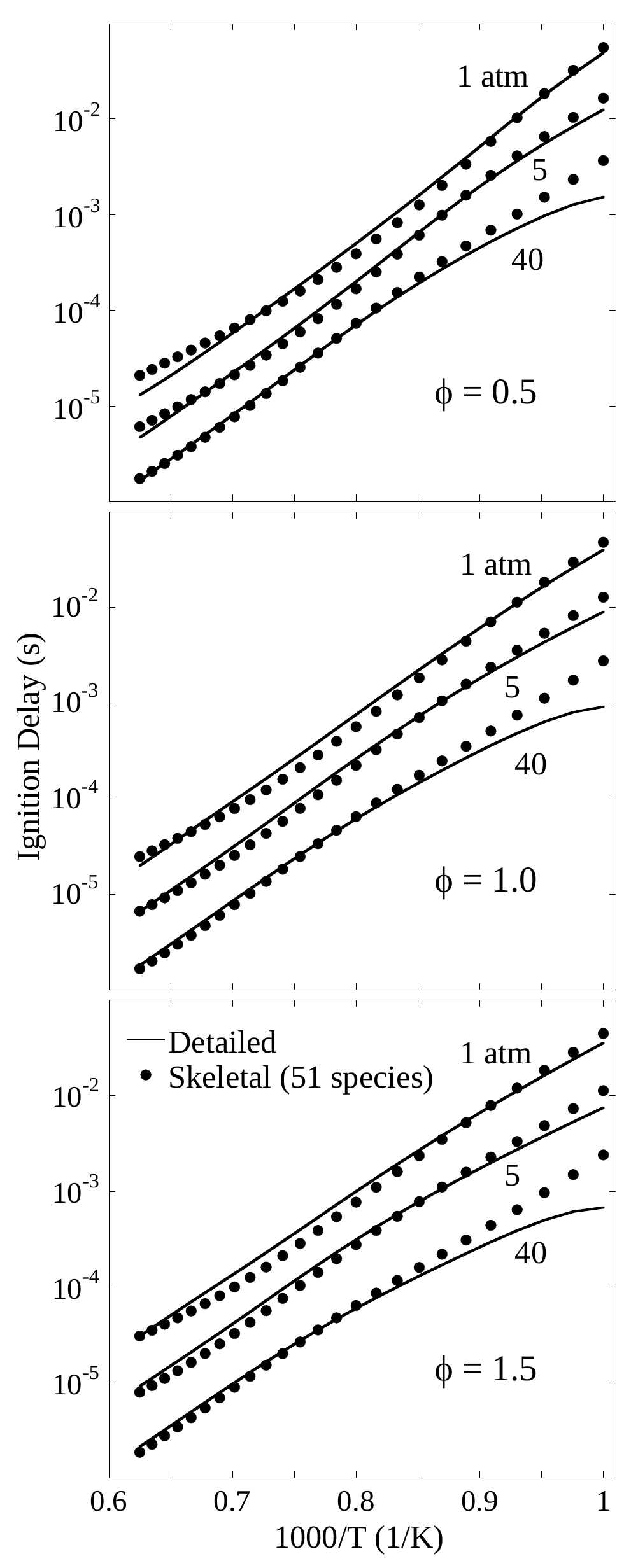}
	\caption{Autoignition validation of \emph{n}-decane high-temperature (51 species and 256 reactions) skeletal mechanism over a range of initial temperatures and pressures, and at varying equivalence ratios.}
	\label{F:ndec-hightemp-results}
\end{figure}

Validation was performed using constant volume autoignition for both the comprehensive and high-temperature skeletal mechanisms covering pressures of 1, 5, and 40 atm and equivalence ratios of 0.5--1.5, with temperature ranges of 600--1600 K and 1000--1600 K shown in Figs.\ \ref{F:ndec-comp-results} and \ref{F:ndec-hightemp-results} respectively.  The comprehensive mechanism well predicts the ignition delay compared to the detailed mechanism for the full range of validation conditions, with some discrepancy in the NTC region.  The high-temperature mechanism also shows fairly good performance, with noticeable discrepancies primarily in the low-temperature, high-pressure region, as expected.

\begin{figure}[htbp]
	\centering
	\includegraphics[width=0.5\linewidth]{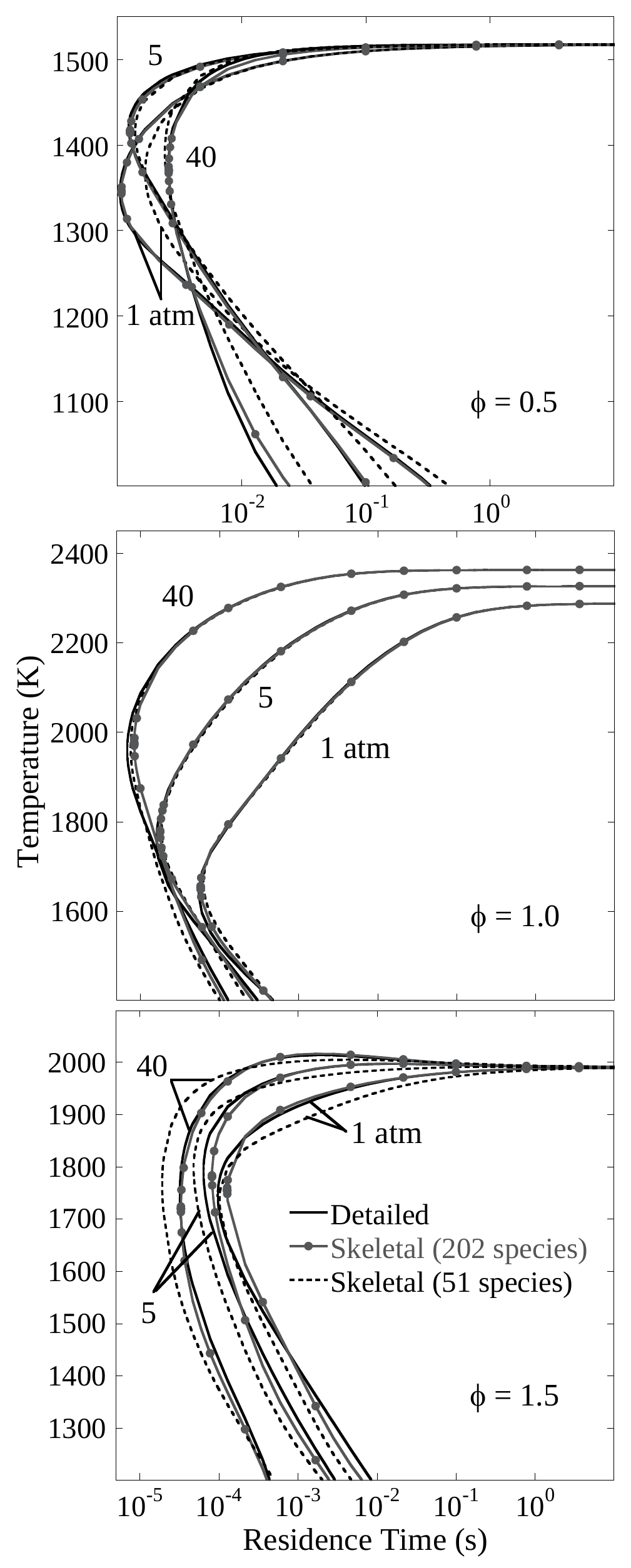}
	\caption{PSR validation of comprehensive (202 species and 846 reactions) and high temperature (51 species and 256 reactions) \emph{n}-decane skeletal mechanisms over a range of residence times and at varying pressures and equivalence ratios with an inlet temperature of 300 K.}
	\label{F:ndec-psr}
\end{figure}

\begin{figure}[htbp]
	\centering
	\includegraphics[width=0.75\linewidth]{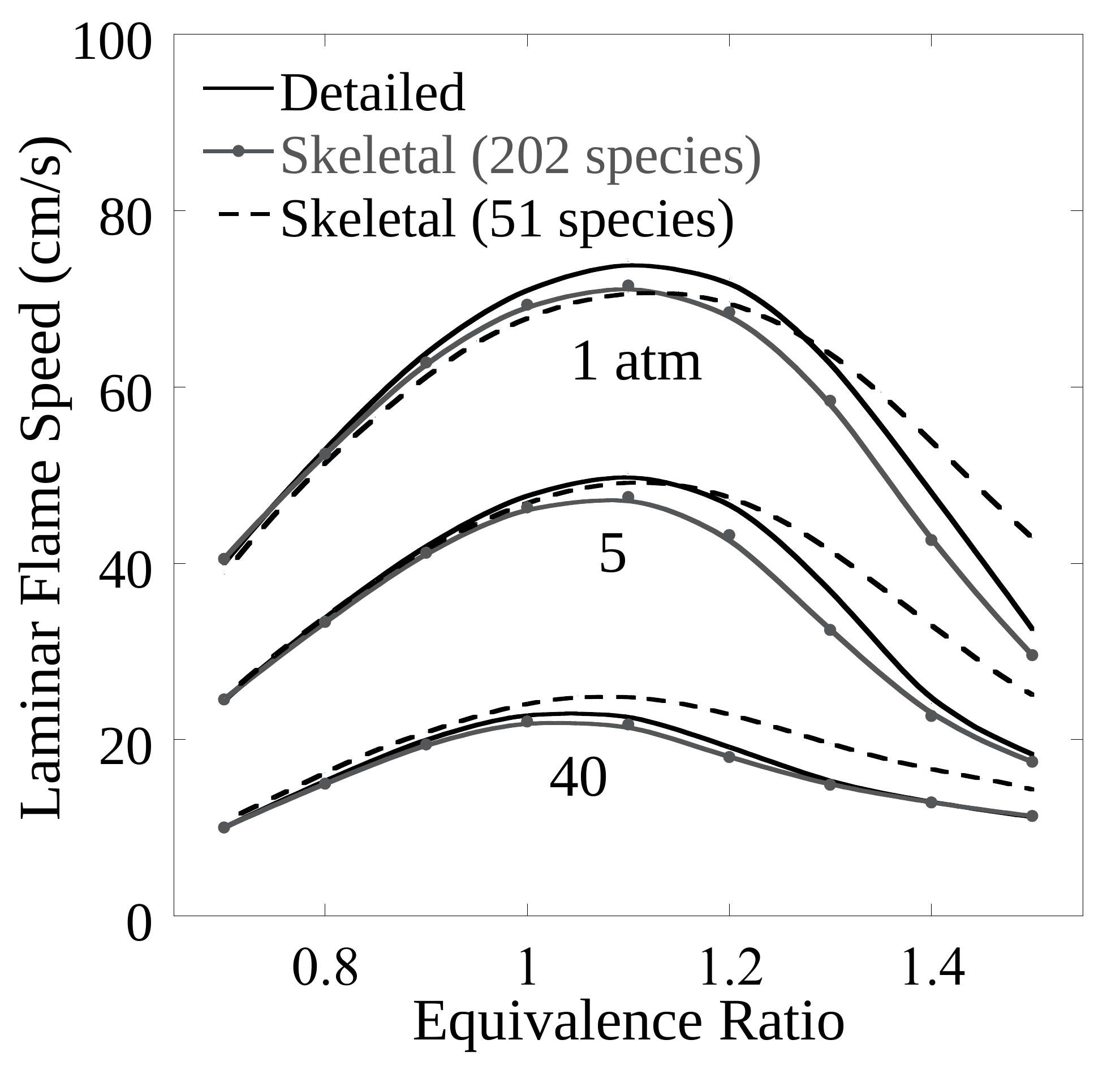}
	\caption{Laminar flame speed validation of comprehensive (202 species and 846 reactions) and high temperature (51 species and 256 reactions) \emph{n}-decane skeletal mechanisms over a range of equivalence ratio and pressure conditions with an unburned mixture temperature of 400 K.}
	\label{F:ndec-premix}
\end{figure}

Both skeletal mechanisms were also validated independently using PSR \cite{Glarborg:1986} and laminar flame speed simulations through PREMIX \cite{Kee:1985}, phenomena not employed to generate kinetics data in the reduction procedure.  Figure \ref{F:ndec-psr} shows the validation of the comprehensive and high-temperature mechanisms in PSR with an inlet temperature of 300 K over pressures of 1, 5, and 40 atm, equivalence ratios of 0.5--1.5, and a range of residence times.  The comprehensive mechanism reproduces the curves, with minor discrepancies only in the rich cases.  The high-temperature mechanism also performs well here, with some discrepancies near the extinction turning points.  Figure \ref{F:ndec-premix} shows the validation in laminar flame speed calculations with an unburned mixture temperature of 400 K and pressures of 1, 5, and 40 atm over a range of equivalence ratios.  The comprehensive skeletal mechanism performs quite well for all cases, with a maximum error of 12.3\% at 5 atm, $\phi = 1.3$.  The high-temperature mechanism also performs well here, with slight discrepancies in laminar flame speed predictions at all equivalence ratios at atmospheric pressure and larger error for the higher pressure cases at rich conditions.

\section{Concluding remarks}
\label{S:conclusion}

In the present work the directed relation graph with error propagation and sensitivity analysis (DRGEPSA) method for skeletal mechanism reduction was presented and discussed.  This approach, a combination of the DRGEP and DRGASA methods, utilizes the specific strengths of each individual method to diminish some of the weaknesses of each.  DRGEP efficiently identifies and removes unimportant species while DRGASA incorporates sensitivity analysis to identify further unimportant species for removal at a greater computational expense.  By combining the two methods, DRGEPSA is able to identify and remove more unimportant species than its precursors.  In addition, the current implementation uses a limited number of user inputs to automatically generate optimally small skeletal mechanisms.  An iterative error threshold selection procedure produces an optimal DRGEP skeletal mechanism for the given error limit before the sensitivity analysis (SA) phase further eliminates unimportant species.  Though the DRGEP phase eliminates a larger number of species in general than DRG, the SA phase must run autoignition simulations for each of the limbo species.  A large list of limbo species combined with a wide range of autoignition conditions under consideration could induce significant computational cost.

Skeletal mechanisms of \emph{n}-heptane were generated to illustrate the improvement of DRGEPSA over DRGEP and DRGASA, resulting in a final mechanism with 108 species compared to 173 and 153 species, respectively.  Skeletal mechanisms of \emph{iso}-octane were also presented to further illustrate the improvement of DRGEPSA over DRGEP and DRGASA, with final skeletal mechanisms of 165, 232, and 211 species, respectively.  All skeletal mechanisms exhibited good ignition delay prediction compared to the detailed mechanisms, with the most noticeable discrepancies in the NTC regions.

Two skeletal mechanisms for \emph{n}-decane were generated using DRGEPSA from a large detailed mechanism for \emph{n}-alkanes, covering \emph{n}-octane through \emph{n}-hexadecane.  One skeletal mechanism covers a comprehensive set of temperature conditions at low to high pressure while the other mechanism is limited to high-temperature conditions, and both mechanisms covered lean to rich equivalence ratios.  The resulting comprehensive skeletal mechanism consists of 202 species and 846 reactions while the high-temperature mechanism is much smaller with 51 species and 256 reactions.  The large extent of reduction for both mechanisms illustrates the capability of the DRGEPSA method to reduce large mechanisms of surrogate fuels.  Both the comprehensive and high-temperature skeletal mechanisms are included as supplementary material to this article.

External validation of the \emph{n}-decane skeletal mechanisms was performed using perfectly-stirred reactor (PSR) and laminar flame simulations.  The comprehensive skeletal mechanism reproduced the results of the detailed mechanism in PSR with larger errors at higher pressure and rich conditions.  The high-temperature skeletal mechanism also performed well, with larger errors at the lean, atmospheric pressure and rich, 40 atm pressure conditions.  The comprehensive skeletal mechanism also performed quite well in predicting laminar flame speed with noticeable error at rich, higher pressure conditions.  By comparison, the high-temperature skeletal mechanism fared slightly less better.  Though only ignition chemical kinetics data were used in the mechanism reduction procedure, both mechanisms performed well predicting the extinction turning point in PSR and even for predicting laminar flame speed where transport phenomena are considered.  It is further noted that if a posteriori validation is not satisfactory, the range of OIC values used to identify limbo species could be adjusted.  One possible pitfall is the removal of certain important species with small induced error in ignition delay but greater importance in other combustion phenomena.  Alternatively, PSR and/or PREMIX simulations could be included in the chemical kinetics data sampling and for error evaluation.

While a significant reduction is achieved with the comprehensive skeletal mechanism (approximately 10\% of the detailed mechanism) using DRGEPSA, the final mechanism is still too large to be used in full-scale three-dimensional simulations.  Nagy and Tur\'{a}nyi \cite{Nagy:2009} suggested that removal of additional unimportant reactions could significantly improve the computational cost of simulations.  An integrated reduction approach involving unimportant reaction removal, isomer lumping, time-scale reduction (e.g.,\ quasi-steady-state assumption), diffusive species bundling, and other reduction methods, similar to the approach presented by Lu and Law \cite{Lu:2008,Lu:2009}, is required before realistic computational simulations are feasible with such skeletal mechanisms; this will be the subject of future work.  However, the small size of the high-temperature skeletal mechanism illustrates the significant reduction capability when the input conditions are limited to the desired range.  For instance, flame simulations rely largely on high-temperature chemistry such that a skeletal mechanism desired for this purpose could omit the complex low-temperature and NTC regime chemistry with acceptable error.  The current high-temperature skeletal mechanism with 51 species and 256 reactions could be used without further reduction in large-scale simulations.

\section*{Acknowledgments}
This work has been supported by the National Aeronautics and Space Administration under grant number NNX07AB36Z, with the technical monitoring of Dr.\ K.\ P.\ Kundu, and the Department of Defense through the National Defense Science and Engineering Graduate (NDSEG) Fellowship program.

\bibliography{refs}
\bibliographystyle{elsarticle-num-CNF}

\pagebreak


\begin{table}[h!]
\begin{center}
\begin{tabular}{l c c c}
\toprule
Method & \# Species & \# Reactions & Max. Error \\
\midrule
DRG & 211 & 1044 & 21\% \\
DRGASA & 153 & 691 & 24\% \\
DRGEP & 173 & 868 & 28\% \\
DRGEPSA & 108 & 406 & 27\% \\
\bottomrule
\end{tabular}
\caption{Comparison of \emph{n}-heptane skeletal mechanism sizes generated by DRG, DRGEP, DRGASA, and DRGEPSA methods.}
\label{T:n-hep1}
\end{center}
\end{table}

\begin{table}[h!]
\begin{center}
\begin{tabular}{l c c c}
\toprule
Method & \# Retained & \# Limbo & \# Removed from Limbo \\
\midrule
DRGASA & 129 & 82 & 58 \\
DRGEPSA & 54 & 119 & 65 \\
\bottomrule
\end{tabular} 
\caption{Comparison of sensitivity analysis results using DRGASA and DRGEPSA methods.}
\label{T:n-hep2}
\end{center}
\end{table}

\begin{table}[h!]
\begin{center}
\begin{tabular}{l c c c}
\toprule
Method & \# Species & \# Reactions & Max. Error \\
\midrule
DRG & 275 & 722 & 13\% \\
DRGASA & 211 & 885 & 26\% \\
DRGEP & 232 & 1140 & 15\% \\
DRGEPSA & 165 & 779 & 19\% \\
\bottomrule
\end{tabular}
\caption{Comparison of \emph{iso}-octane skeletal mechanism sizes generated by DRG, DRGEP, DRGASA, and DRGEPSA methods.}
\label{T:iso-oct}
\end{center}
\end{table}

\pagebreak

\listoffigures

\pagebreak

\end{document}